\documentstyle[12pt,epsf]{article}
\newcommand{\be}{\begin{equation}}
\newcommand{\ee}{\end{equation}}
\newcommand{\lb}[1]{\label{#1}}
\begin{document}
\font \gothic=eufm10  scaled \magstep1
\font \gothind=eufm7
\newcommand\goth[1]{{\mbox{{\gothic #1}}}}
\begin{center}
\begin{Large}
{\bf{On a possibility of phase transitions in the geometric\\
     structure of space-time}} \footnote{This research is supported
  by DFG grant No.7/25/94 and INTAS--93--1630--EXT grant}\\
\end{Large}   
\vskip 5mm
\begin{large}
      G.Yu. Bogoslovsky \footnote{E-mail: bogoslov@theory.npi.msu.su} \\
\end{large}
{\it{Institute of Nuclear Physics, Moscow State University,
     119899 Moscow, Russia}}\\
\begin{large}
H.F. Goenner \footnote{E-mail: goenner@theorie.physik.uni-goettingen.de} \\
\end{large}
 {\it{Institute for Theoretical Physics, University of G\"ottingen,
     D-37073, G\"ottingen, Germany}}\\

\end{center}
\vskip 7mm
\hrule
\vskip 3mm
\noindent
{\bf{Abstract}}
\vskip 3mm
     It is shown that space-time may be not only in a state which is 
described by Riemann geometry but also in states which are described
by Finsler geometry. Transitions between various metric states of
space-time have the meaning of phase transitions in its geometric
structure. These transitions together with the evolution of each of the
possible metric states make up the general picture of space-time
manifold dynamics.\\
\vskip 3mm
\noindent
{\sl{PACS\,:}}\, 95.30.-k; 95.30.Sf; 98.80.Bp; 98.80.Hw; 02.40.-k 
                 Relativistic cosmology
\vskip 3mm
\hrule
\vskip 2cm

    According to the contemporary cosmological models [1], space-time is
Riemannian, i.e. locally isotropic, and preserves its local isotropy
in the process of the Universe evolution. At the same time there exist
some indirect indications of the fact that nowadays space-time has a weak
relic local anisotropy and therefore it is not unlikely that it is 
described by Finsler geometry [2] rather than by Riemann geometry.    
Moreover, if it appears that the Hubble constant has no dipole anisotropy
correlated with the dipole anisotropy of the microwave background radiation
then it will unequivocally indicate a strong local anisotropy of space-time
at the early stage of the Universe evolution.

   One of the possible mechanisms of the appearance of a local anisotropy
in space-time is the induced phase transition in its geometric structure, 
caused by the breakdown of higher gauge symmetries and by the appearance
of masses in fundamental fields of matter. This involves changes in the
metric properties of space-time manifold and it goes over from a state
described by Riemann geometry into a state described by Finsler geometry.
Since Finslerian space-time differs from Riemannian space-time by the
anisotropy of its tangent spaces, in such a transition there occurs
a flagrant violation of the local Lorentz symmetry of space-time. In the
course of subsequent expansion of the Universe the initial strong local
anisotropy of the Finslerian space-time monotonically decreases and, on
the average, tends to zero together with its curvature. Gradually the
local Lorentz symmetry of space-time is also restored.

   The first indication of the fact that nowadays the local Lorenz symmetry
still remains slightly broken was obtained from the investigation of the 
spectrum of primary cosmic superhigh-energy protons. The point is that
according to the calculations [3,\,4], which substantially employ the local
Lorentz symmetry of space-time, the proton energy spectrum should be cut
off (\,due to the intense production of pions on relic radiation photons\,) at
proton energies $\sim 5\times 10^{19} eV.$ The experimental data [5,\,6], 
however, are most likely indicative of the absence of such an effect. This
situation induced the investigators [7,\,8] to assume that the conventional
Lorentz transformations become invalid for the Lorentz factors
$\gamma >5\times 10^{10} $ and the correct relation between the various
inertial reference frames at any values of $\gamma$ is provided by the
other, so-called generalized Lorentz transformations. Subsequently [9] 
it has indeed become possible to find the generalized Lorentz
transformations. It appeared that they belong to a group of local
relativistic symmetry of Finslerian space-time, in which case the smaller
the local anisotropy of Finslerian space-time, i.e. the closer is it to
Riemannian one, the closer to the velocity of light tend the generalized
Lorentz transformations to be markedly different from the conventional
ones. Therefore the use of these transformations in calculating the cutoff
point of the primary cosmic proton spectrum enables one, in principle, to
remove the emerged discrepancy between the theoretical predictions and the
experimental data pertaining to the superhigh energy region.

   In order to demonstrate that the existence of the generalized Lorentz 
transformations necessarily leads to the existence of a local anisotropy
in space-time, first consider a two-dimensional event space. In this case
the generalized Lorentz transformations appear as
\be\lb{1}
\left\{
\begin{array}{lclll}
   x'_0&=&e^{-r\alpha}&(&x_0\cosh\alpha -x\sinh\alpha)\\
   x'&=&e^{-r\alpha}&(-&x_0\sinh\alpha +x\cosh\alpha),
\end{array}
\right.
\ee
where\, $ \tanh \alpha =v/c $ \,and\, $r$ \,is the dimensionless parameter of the
scale transformation. It is obvious that in spite of the presence of
additional dilatation the transformations (1) still remain linear,
constitute a group with the group parameter $\alpha $ and lead to
Einstein's law of velocity addition. However, the pseudo-Euclidean metric
is no longer their invariant. It is easy to verify that the invariant of
the transformations (1) is the metric
\be\lb{2}
ds^2=\left[\frac{(dx_0-dx)^2}{dx_0^2-dx^2}\right]^r(dx_0^2-dx^2)\,.
\ee
The given metric belongs to a class of Finslerian metrics and describes
a flat but anisotropic space of events. While we consider a two-dimensional
anisotropic space, its anisotropy manifests itself in noninvariance of the
metric (2) under the reflection transformation
 $x_0\to -x_0$  or  $x\to -x\,.$ The anisotropy just mentioned disappears
only in the case $r=0\,,$ when the event space becomes pseudo-Euclidean
and the generalized Lorentz transformations (1) become the conventional
Lorentz ones. Therefore the parameter $r$ characterizes the value of the
space anisotropy.

   If in (2) we replace the forms  $(dx_0^2-dx^2)$ and $(dx_0-dx)$ by
their four-dimensional analogs, i.e. make the substitution
$$
(dx_0^2-dx^2)\rightarrow (dx_0^2-d\vec x^{\,2})\,;\quad 
(dx_0-dx)\rightarrow (dx_0-\vec\nu d\vec x)\,,
$$
then we can thus obtain the corresponding four-dimensional metric
\be\lb{3}
ds^2=\left[\frac{(dx_0-\vec\nu d\vec x)^2}{dx_0^2-d\vec x^{\,2}}\right]^r
(dx_0^2-d\vec x^{\,2})\,.
\ee

     The Finslerian metric (3) describes a flat anisotropic space-time
with partially broken rotational symmetry. This signifies that instead of
the 3-parameter rotation group, which was admitted by the isotropic 
pseudo-Euclidean event space, the space-time (3) admits only a 1-parameter
group of rotations about the unit vector\, $\vec\nu $ \,which indicates the
preferred direction in a 3-space. As a result the homogeneous isometry
group of the space-time (3) turns out to be a 4-parameter group [9] rather
than a 6-parameter group, in which case as the transformations of
relativistic symmetry it incorporates a 3-parameter noncompact subgroup
of the generalized Lorentz transformations
\be\lb{4}
x'^i=D(\vec v,\vec\nu )\,R^i_j(\vec v,\vec\nu )\,L^j_k(\vec v)\,x^k\,,
\ee
where $L^j_k(\vec v)$ is a conventional Lorentz boost, 
$R^i_j(\vec v,\vec\nu )$ is the rotation of the space axes about the vector
$[\vec v\,\vec\nu ]$ through an angle determined by relativistic aberration
of the preferred direction 
$\vec\nu\,,\, 
\ D(\vec v,\vec\nu)=\left[(1-\vec v\vec\nu /c)/\sqrt{1-\vec v^{\,2}/c^2}
\right]^rI\,,$ 
and 
$I$ 
is the unit matrix.

     The difference of the metric (3) from the pseudo-Euclidean one is such
that it does not involve the light cone equation. Therefore the 3-geometry
still remains Euclidean. At the same time the anisotropy of the event space
(3) leads to nontrivial consequences even at the level of nonrelativistic
physics. In particular [10]\,, the effective inertness of a particle of
mass $m$ turns out to be dependent on the quantities $r$ and $\vec\nu \,,$
which characterize space anisotropy, and is determined by a tensor of
inert mass 
\be \lb{5} 
{\goth M}_{\alpha\beta}=m(1-r)({\delta}_{\alpha\beta}+r{\nu}_\alpha\,{\nu}_
\beta )\,.  
\ee
Thus Newton's second law takes the form\,
${\goth M}_{\alpha\beta}\,a^\beta =F_\alpha \,.$

     It should be noted here that the parameters $r$ and $\vec\nu $ are
in fact local values of the corresponding fields $r\,(\,x\,)$ and
$\nu _i\,(\,x\,)\,,$ in which case $\nu _i\,\nu ^i=0\,.$ Together with
the field of the Riemannian metric tensor $g_{ik}\,(\,x\,)\,,$ responsible
for gravitation, these fields determine the Finslerian metric of a curved
locally anisotropic space-time 
\be \lb{6}
ds^2 = \left [\frac{(\,\nu _i\,dx^i\,)^2}{g_{ik}\,dx^idx^k}\right ]^{r}\,
(\,g_{ik}\,dx^idx^k\,)\,.
\ee
This space-time has flat anisotropic spaces (3) as tangent spaces,
possesses the 3-parameter group of local relativistic symmetry (4) and
reduces to the Riemannian space-time at $r\,(\,x\,)=0\,.$ Noteworthy is
also the fact that the dynamics of the Finsler space (6) is completely
determined by the dynamics of the fields $g_{ik}\,(\,x\,)\,, r\,(\,x\,)$
and $\nu _i\,(\,x\,)$ and is described by a system of field equations
[11-13] generalizing the Einstein equations. In this case, within the
framework of the Finslerian theory of gravitation, the particle inert
mass turns out, according to (5), to be a tensor field on space-time
and is ultimately determined (\,in accordance with the Mach principle\,) 
by the distribution and motion of external matter. The possibility of
realizing the Mach principle is characteristic of the Finslerian theory.
In the metric theories of gravitation, employing the Riemannian model
of space-time, there is no such possibility since there is no local
anisotropy field itself, i.e. $r\,(\,x\,)=0\,.$
  
     Obviously, the maximum permissible local value of the field 
$r\,(\,x\,)$ is a value of $r=1\,.$ At such $r$ the linear element 
of the tangent space (3) degenerates into the total differential
\be \lb {7}
ds=dx_0-\vec\nu \,d\vec x
\ee
and, consequently, the action $S=-mc\int_a^b\,ds$ for a free particle
of mass $m$ is no longer dependent on the shape of the world line connecting
the points $a$ and $b\,.$ This means that at $r=1$ any massive particle
loses its inertness. The aforesaid is illustrated by (5) in accordance
with which ${\goth M}_{\alpha \beta }=0$ at $r=1\,.$ At $r=1\,,$ along
with inertness the notion of spatial extension disappeares, which is due
to the absence of a light cone in this case and, consequently, of the
possibility itself of determining spatial distances using the exchange
of light signals. As a result, in the space-time (7) there remains the
single physical characteristic, namely, time duration and it should be
regarded as an interval of absolute time.
 
   Since the "metric" (7) is a special case of the metric (3), then to
within isomorphism all the transformations, which leave invariant the
metric (3), leave invariant the "metric" (7) as well. At the same time,
on making the substitution of the variables
$\nu _1\,x_1 \to x_1\,, \,\nu _2\,x_2 \to x_2\,, \,\nu _3\,x_3 \to
x_3\,; \,\,\nu _1\,, \nu _2\,, \nu _3\,\ne 0\,$ and on representing (7) as
\be \lb{8}
ds=dx_0-dx_1-dx_2-dx_3\,,
\ee
one can find that in comparison with (3) the "metric" (7) has an additional
Abelian 3-parameter isometry group. It turned out [14] that there exists
a homogeneous noncompact group which to within isomorphism coincides with
the above-mentioned Abelian group and is a group of relativistic symmetry
of a flat space-time with the metric
\be \lb{9}
\begin{array}{rl}
ds=(dx_0-dx_1-dx_2-dx_3)^{(1+r_1+r_2+r_3)/4}(dx_0-dx_1+dx_2+dx_3)^
{(1+r_1-r_2-r_3)/4}&\\
\times \ (dx_0+dx_1-dx_2+dx_3)^{(1-r_1+r_2-r_3)/4}(dx_0+dx_1+dx_2-dx_3)^
{(1-r_1-r_2+r_3)/4}&. 
\end{array}             
\ee
The given Finslerian metric depends on three parameters $r_1\,,r_2$ 
and $r_3$ and describes an anisotropic space-time with the entirely
broken symmetry with respect to the rotation group. The permissible
values of the parameters\, $r_1\,,r_2$ \,and\, $r_3$ \,are limited by the
conditions
$$
\begin{array}{ll}
1+r_1+r_2+r_3>0\,,&1+r_1-r_2-r_3>0\,,\\
1-r_1+r_2-r_3>0\,,&1-r_1-r_2+r_3>0\,,\\
\end{array}
$$
which ensure the fact that the section of a light cone by hyperplane
$dx_0=const$ is a closed convex surface, and this ensures the 
applicability of the procedure of exchange of light signals for
determining 3-space distances.
 
    In the limiting case, where, for example, $r_1=r_2=r_3=1\,,$
the metric (9) degenerates into a 1-form (8), i.e. into the total
differential of absolute time. If we now recall that the metric 
(3) of the flat anisotropic space-time with the partially broken
3-rotational symmetry also degenerates at $r=1$ into the total
differential of absolute time, then it suggests that absolute time
is not a stable degenerate state of space-time and as a result of
the "primary" phase transition it may turn either into the 
partially anisotropic space-time (3) or into the entirely anisotropic
space-time (9)\,. In any case such a phase transition is accompanied
by an "act of creation" of a three-dimensional space, in which case
its geometry depends on the direction of the phase transition. In
the passage to (3) there occurs a 3-space with locally Euclidean
geometry while in the passage to (9) there occurs, as will be shown
below, a 3-space with locally non-Euclidean geometry. Thus it is 
precisely absolute time that is a connecting link by which the 
principle of  correspondence is satisfied for the Finsler spaces
(3) and (9)\,.
  
     It has already been pointed out that the homogeneous isometry 
group of the event space (9) is an Abelian 3-parameter noncompact
group. The transformations belonging to it link various inertial
reference frames and are of the form
\be \lb{10}
x'_i=D\,L_{ik}\,x_k\,,
\ee
where: $\ D=\exp (-r_1\,\alpha _1-r_2\,\alpha _2-r_3\,\alpha _3\,)\,;$
\ the matrices
$$
L_{ik}=\left (
\begin{array}{rrrr}
\cal A&-\cal B&-\cal C&-\cal D\\
-\cal B&\cal A&\cal D&\cal C\\
-\cal C&\cal D&\cal A&\cal B\\
-\cal D&\cal C&\cal B&\cal A\\
\end{array}
\right )
$$
are unimodular, whereby
$$
{\cal A}=\cosh \alpha _1\cosh \alpha _2\cosh \alpha _3+
\sinh \alpha _1\sinh \alpha _2\sinh \alpha _3\,,
$$
$$
{\cal B}=\cosh \alpha _1\sinh \alpha _2\sinh \alpha _3+
\sinh \alpha _1\cosh \alpha _2\cosh \alpha _3\,,
$$
$$
{\cal C}=\cosh \alpha _1\sinh \alpha _2\cosh \alpha _3+
\sinh \alpha _1\cosh \alpha _2\sinh \alpha _3\,,
$$
$$
{\cal D}=\cosh \alpha _1\cosh \alpha _2\sinh \alpha _3+
\sinh \alpha _1\sinh \alpha _2\cosh \alpha _3\,;
$$
and $\alpha _1\,,\alpha _2\,,$ and $\alpha _3$ are group parameters.
   
     In place of $\alpha _1\,,\alpha _2\,,$ and $\alpha _3$ as group
parameters we can use the velocity components $\,v_1\,,v_2\,,$ and
$\,v_3\,$ of the primed reference frame. For this purpose it is 
sufficient to put $\,x'_1=x'_2=x'_3=0\,$ in the transformations inverse
to (10)\,. As a result we arrive at the relations
$$v_1=(\tanh\alpha _1-\tanh\alpha_2\tanh\alpha _3)/(
1-\tanh\alpha_1\tanh\alpha_2\tanh\alpha_3)\,,
$$
$$v_2=(\tanh\alpha _2-\tanh\alpha_1\tanh\alpha _3)/(
1-\tanh\alpha_1\tanh\alpha_2\tanh\alpha_3)\,,
$$
$$v_3=(\tanh\alpha _3-\tanh\alpha_1\tanh\alpha _2)/(
1-\tanh\alpha_1\tanh\alpha_2\tanh\alpha_3)\,.
$$
\vspace{2mm}
\noindent
The inverse relations appear as
\vspace{2mm}
$$
\alpha _1\,=\frac{1}{4}\ln \frac{(1+v_1-v_2+v_3)(1+v_1+v_2-v_3)}
{(1-v_1-v_2-v_3)(1-v_1+v_2+v_3)}\,,
$$
$$
\alpha _2\,=\frac{1}{4}\ln \frac{(1-v_1+v_2+v_3)(1+v_1+v_2-v_3)}
{(1-v_1-v_2-v_3)(1+v_1-v_2+v_3)}\,,
$$
$$
\alpha _3\,=\frac{1}{4}\ln \frac{(1-v_1+v_2+v_3)(1+v_1-v_2+v_3)}
{(1-v_1-v_2-v_3)(1+v_1+v_2-v_3)}\,.
$$
\vspace{2mm}
 
    Noteworthy is the fact that such an observable as, for example,
the velocity $v$ is no longer determined now by the formula
$\,v=\sqrt{v_1^2+v_2^2+v_3^2}\,.$ This comes from the fact that
in the case of the flat event space (9) the geometry of the 
corresponding 3-space turns out to be non-Euclidean while time
$x_0\,,$ to be coordinate time only.
   
    In order to determine how the difference of coordinates of
two events in the event space (9) is related to observables and
thereby to obtain correct formulas for the observables it is 
necessary to use a procedure involving the exchange of light 
signals between neighbouring points of the 3-space. For this
purpose we first of all note that according to the definition (9)
the tolerance range of $\,dx_i\,$ values is limited by the conditions
\be \lb{11}
\left\{
\begin{array}{rcl}
dx_0-dx_1-dx_2-dx_3&\ge&0\\
dx_0-dx_1+dx_2+dx_3&\ge&0\\
dx_0+dx_1-dx_2+dx_3&\ge&0\\
dx_0+dx_1+dx_2-dx_3&\ge&0\,.\\
\end{array}
\right.
\ee
These conditions determine either the timelike interval between
two events or the interval equal to zero and are invariant under
the relativistic transformations (10)\,. Apart from this the
transformations (10) leave invariant the sign of $\,dx_0\,.$ Let 
now $\,dx_0>0\,.$ Then in terms of the component $\,v_\alpha
=dx_\alpha\,/dx_0\,$ of the coordinate velocity the conditions (11) 
can be rewritten as follows
$$
\left\{
\begin{array}{rcl}
1-v_1-v_2-v_3&\ge&0\\
1-v_1+v_2+v_3&\ge&0\\
1+v_1-v_2+v_3&\ge&0\\
1+v_1+v_2-v_3&\ge&0\,.\\
\end{array}
\right.
$$
The range of $\,v_\alpha \,$ values, limited by these conditions,
is depicted in Figure 1\,. The range considered represents a regular
tetrahedron with the center at the origin $\,o\,$ of a rectangular
system of the coordinates $\,v_1\,,v_2\,,v_3\,.$ The velocities
corresponding to the timelike intervals $\,ds\,$ fill the inner
region of the tetrahedron while the velocities, which describe the
propagation of light signals and ensure the fulfillment of the
equality $\,ds=0\,,$ fill the surface of the tetrahedron. Figure 1,
next to their letterings, gives the coordinates of eight of
the fourteen characteristic points which lie on the tetrahedron
surface. The coordinates of the remaining six points are as follows
 $\,\varepsilon \,(0\,,-1\,,0)\,;\,\, \beta \,(0\,,1\,,0)\,;\,\, 
 \Gamma \,(1/3\,,1/3\,,1/3)\,;\,\, \Theta \,(-1/3\,,-1/3\,,1/3)\,;\,\, 
 \Phi \,(-1/3\,,1/3\,,-1/3)\,;\\ \Omega \,(1/3\,,-1/3\,,-1/3)\,.$ 
The role of these points consists in that with the aid of them
the tetrahedron surface is divided into 12 equal tetragons and they,
in turn, are grouped into six pairs of mutually conjugated (\,with
respect to a reflection operation at the origin\,) tetragons. By
labelling the reflection operation with a 
symbol \  $\longleftrightarrow $ \  we obtain the following pairs
 
$$
\begin{array}{lcl}
\Gamma \gamma \Delta \beta &\ \longleftrightarrow \ &\Psi \zeta \Omega
\varepsilon \\
\Gamma \beta \Upsilon \alpha &\ \longleftrightarrow \ &\Psi \varepsilon
\Theta \delta \\
\Gamma \alpha \Lambda \gamma &\ \longleftrightarrow \ &\Psi \delta
\Phi \zeta \\
\Omega \varepsilon \Lambda \alpha &\ \longleftrightarrow \ &\Delta
\beta \Phi \delta \\
\Omega \alpha \Upsilon \zeta &\ \longleftrightarrow \ &\Delta \delta
\Theta \gamma \\
\Theta \gamma \Lambda \varepsilon &\ \longleftrightarrow \ &\Upsilon
\zeta \Phi \beta \,.\\
\end{array}
$$

Consider, for example, a tetragon $\,\Gamma \gamma \Delta \beta \,.$ The 
points filling it determine the coordinate velocities of light signals
propagating within a solid angle (\,sector\,) $\Gamma \gamma \Delta \beta
o\,.$ Rays propagating in the opposite directions belong to the sector 
$\,\Psi \zeta \Omega \varepsilon o\,$ while the velocities of light,
corresponding to them, fill the tetragon $\,\Psi \zeta \Omega \varepsilon
\,.$ If $\,(\,v_1\,,v_2\,,v_3\,)\,$ are the velocity components of an
initial light signal propagating in the sector $\,\Gamma \gamma \Delta \beta
o\,$ and $\,(\,{\tilde v}_1\,,{\tilde v}_2\,,{\tilde v}_3\,)\,$ are the 
velocity components of the reflected signal, i.e. the signal belonging to
the sector $\,\Psi \zeta \Omega \varepsilon o\,,$ then, as can readily be
verified,
\be \lb{12}
{\tilde v}_1=-\frac{v_1}{v_2+v_3-v_1}\,;\ {\tilde v}_2=
-\frac{v_2}{v_2+v_3-v_1}\,;\ {\tilde v}_3=-\frac{v_3}{v_2+v_3-v_1}\,,
\ee 
   
    Let now $\,(\,0\,,0\,,0\,,0\,)\,$ be the coordinates of the event
associated with the emission of a light signal within the 
sector $\,\Gamma \gamma \Delta \beta o\,$ and
$\,(\,dx_0^{(1)}\,,dx_1\,,dx_2\,,dx_3\,)\,$ are the coordinates of the event
associated with the reflection of the given signal. In this case 
$\,(\,dx_0^{(1)}+dx_0^{(2)}\,,0\,,0\,,0\,)\,$ are the coordinates of the
event involving the return of the signal to the initial point. Then the
coordinate velocities of the initial and reflected signals can be represented,
respectively, as
\be \lb{13}
v_1={dx_1}/{dx_0^{(1)}}\,,\, v_2={dx_2}/{dx_0^{(1)}}\,,\,
v_3={dx_3}/{dx_0^{(1)}}\,\,;
\ee
\be \lb{14}
{\tilde v}_1=-{dx_1}/{dx_0^{(2)}}\,,\,{\tilde v}_2=-
{dx_2}/{dx_0^{(2)}}\,,\,
{\tilde v}_3=-{dx_3}/{dx_0^{(2)}}\,\,.
\ee
It follows from (13),\,(14) and (12) that
\be \lb{15}
dx_0^{(2)}/dx_0^{(1)}=v_2+v_3-v_1\,\,.
\ee
Since the initial ray is prescribed within the sector $\,\Gamma \gamma \Delta
\beta o\,$ and the tetragon $\,\Gamma \gamma \Delta \beta \,$ belongs to the
facet $\,\Delta \Upsilon \Lambda \,$ then the components $\,v_{\alpha }\,$ 
satisfy the equation
\be \lb{16}
1-v_1-v_2-v_3=0\,.
\ee
Using (15) and (16) and taking into account the definition (13), we thus
arrive at the relations
\be \lb{17}
(\,dx_0^{(1)}+dx_0^{(2)}\,)/2=dx_2+dx_3\,,
\ee
\be \lb{18}
(\,dx_0^{(1)}-dx_0^{(2)}\,)/2=dx_1\,.
\ee
By its meaning formula (17) determines within the sector 
$\,\Gamma \gamma \Delta \beta o\,$ the 3-space distance between the points 
$\,(\,0\,,0\,,0\,)\,$ and $\,(\,dx_1\,,dx_2\,,dx_3\,)\,,$ that is, it
 determines the 3-metric in the form
\be \lb{19}
dl=dx_2+dx_3\,.
\ee
  
     According to the definition of the coordinates of events, at the 
instant of signal reflection a coordinate clock, placed at the reflection
point, indicates the time $\,dx_0^{(1)}\,$ and a coordinate clock, placed
at the point of signal emission, indicates the time 
$\,(\,dx_0^{(1)}+dx_0^{(2)}\,)/2\,.$ These indications differ by the value
$$
\Delta x_0=dx_0^{(1)}-\frac{dx_0^{(1)}+dx_0^{(2)}}{2}=
\frac{dx_0^{(1)}-dx_0^{(2)}}{2}\,.
$$
Therefore, taking into account the relation (18), the coordinate clocks within
the sector $\,\Gamma \gamma \Delta \beta o\,$ are synchronized using the
following algorithm
\be \lb{20}
\Delta x_0=dx_1\,.
\ee
    
     Finally determine the observable value $\,v\,$ of the particle velocity
within the sector $\,\Gamma \gamma \Delta \beta o\,.$ Assume that at the
instant of starting of a particle from the point with the space coordinates 
$(\,0\,,0\,,0\,)$ the first clock placed at this point indicated the time 
$\,0\,$ while at the instant of its arrival at the point 
$(\,dx_1\,,dx_2\,,dx_3\,)$ the second clock present there indicates the
time $\,dx_0\,.$ Since, according to (20), at the instant of particle
starting the second clock indicated the time $\,\Delta x_0=dx_1\,$ rather
than the time $\,0\,,$ the true time interval spent on the displacement 
$\,d\vec x\,$ is $\,d\tau =(\,dx_0-dx_1\,)\,$ rather than $\,dx_0\,.$ 
Considering now that the length of the vector $\,d\vec x\,$ is calculated
with the aid of (19), we conclude that
\be \lb{21}
v=\frac{dl}{d\tau }=\frac{dx_2+dx_3}{dx_0-dx_1}=\frac{v_2+v_3}{1-v_1}\,.
\ee
According to (21), $\,v\le 1\,,$ in which case for a photon $\,v=1\,.$ In 
the latter case (21) is equivalent to (16) and, hence,  $\,ds=0\,.$
  
     Having made similar calculations we can determine the observables 
 $\,dl\,,\Delta x_0\,$ and $\,v\,$ for each of the remaining eleven 
 sectors. The complete set of the corresponding formulas is presented in 
 the Table of observables. Using this Table we demonstrate what, for
 example, an Euclidean image of the non-Euclidean sphere of radius $\,dl\,$ 
is. For this purpose let us introduce a rectangular system of coordinates 
$\,dx_1\,, dx_2\,, dx_3\,$ in the Euclidean 3-space. One can readily see
that each of the twelve sectors cuts its own piece (\,a rhomb\,) out of the 
corresponding plane $\,dl=const\,.$ All the twelve rhombs turn out to be 
equal to each other and they together make up the surface of a regular 
rhombic dodecahedron. Such a dodecahedron is depicted in Figure 2\,. In 
fact it represents the indicatrix of a flat 3-space whose symmetry is
 determined by a group of the corresponding discrete transformations
 rather than by the rotation group. 
   
      Having considered, along with the locally isotropic Riemannian metric, 
Finslerian metrics with the partially (3) and entirely (9) broken local 
isotropy, as a result we have obtained their unified description. It is 
generally agreed that the evolution of the Universe occurs within the 
framework of the Riemannian geometric model of space-time. At the same 
time our results [15] indicate that space-time may be not only in a state which is 
described by Riemann geometry but also in states which are described
by Finsler geometry. Transitions between various metric states of
space-time have the meaning of phase transitions in its geometric
structure. These transitions together with the evolution of each of the
possible metric states make up the general picture of space-time
manifold dynamics. 
   
     In conclusion we note that apart from the Riemannian metric only two 
types of Finslerian metrics, namely, the metrics (3) and (9), admit 
3-parameter groups of local relativistic symmetry. Therefore within the 
framework of classical theory the three cited metrics make up the complete 
ensemble of the possible metric states of space-time manifold. In the 
corresponding quantum theory this ensemble is substantially expanded to 
admit the inclusion of Finslerian metrics [16] with partially broken 
local relativistic symmetry and also of Finslerian metrics [17,\,18] which 
describe Berwald spaces with entirely broken local relativistic symmetry.

\vskip 6.5cm
 
\noindent
\begin{large}
{\bf{Figure and Table captions}}\\
\end{large} 
   
   Fig.\,1 : The relativistically invariant range of permissible  
   $\,v_\alpha \,$ values.\\   

   Fig.\,2 : A regular rhombic dodecahedron as an Euclidean image
   of the sphere of radius\\
   \phantom{aaa}$dl$\,, prescribed in the
   flat non-Euclidean 3-space.\\
        
   Tabl. : Table of observables.\\

\clearpage
\begin{figure}
\centering
  \unitlength 1cm
 \begin{picture}(16,12.6)
\put(2.7,0.2){\epsfxsize=10.6cm \leavevmode \epsfbox{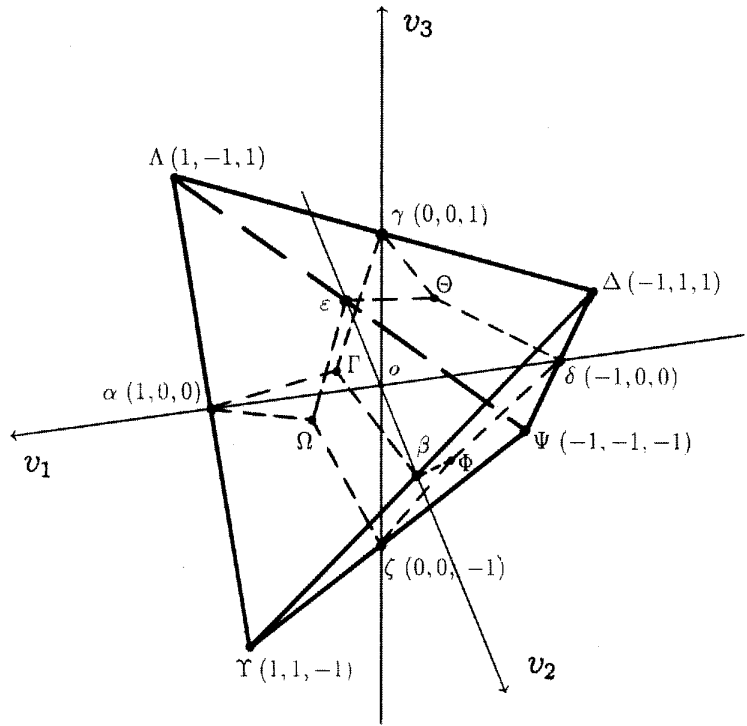}}
\end{picture} 
\caption{The relativistically invariant range 
         of permissible $\,v_\alpha \,$ values.} 
\end{figure} 
\clearpage        
   
\begin{figure}
\centering
\unitlength 1cm
\begin{picture}(16,12.6)
\put(2.7,0.2){\epsfxsize=10.6cm \leavevmode \epsfbox{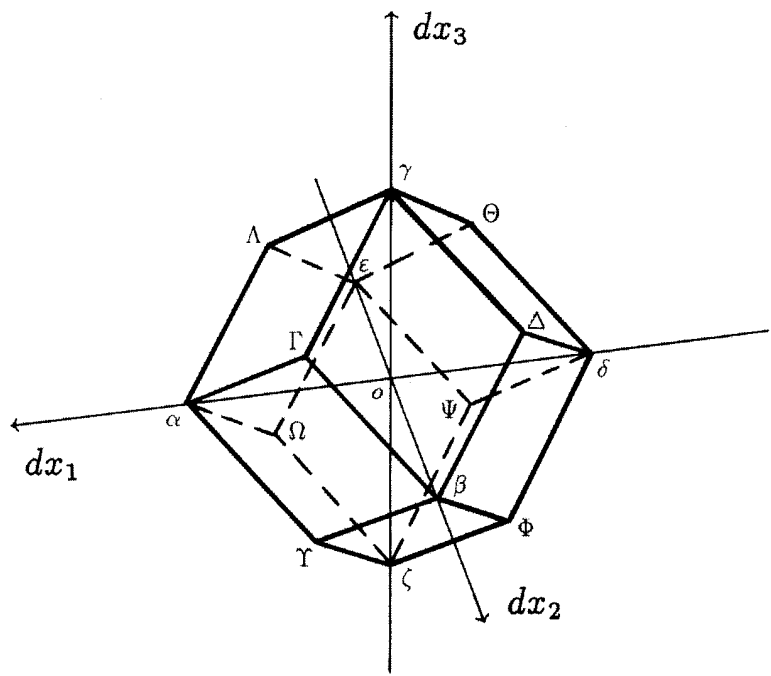} }
\end{picture} 
\caption{A regular rhombic dodecahedron as an 
Euclidean image of the sphere of radius $\,dl\,,$
prescribed in the flat non-Euclidean 3-space.} 
\end{figure}
\clearpage

\vspace*{5cm}

\begin{center}
Table of observables
\vskip 5mm
\begin{tabular}{|r c|c|c|c|}
\hline
\multicolumn{2}{|c}{sector}
    &\multicolumn{1}{|c}{$dl$}
      &\multicolumn{1}{|c}{$\Delta x_0$}
        &\multicolumn{1}{|c |}{$v$}\\
\hline \hline
$\Gamma \gamma\Delta \beta o$ &  & $dx_2+dx_3$
& $dx_1$ & $(v_2+v_3)/(1-v_1)$ \\
\hline
 &$\Psi \zeta \Omega \varepsilon o$ & $-(dx_2+dx_3)$
& $dx_1$ &$-(v_2+v_3)/(1-v_1)$ \\
\hline 
$\Gamma \beta \Upsilon \alpha o$ &  & $dx_1+dx_2$
& $dx_3$ & $(v_1+v_2)/(1-v_3)$ \\
\hline
 &$\Psi \varepsilon \Theta \delta o$ & $-(dx_1+dx_2)$
& $dx_3$ & $-(v_1+v_2)/(1-v_3)$ \\
\hline
$\Gamma \alpha \Lambda \gamma o$ &  & $dx_1+dx_3$
& $dx_2$ & $(v_1+v_3)/(1-v_2)$ \\
\hline
 &$\Psi \delta \Phi \zeta o$ & $-(dx_1+dx_3)$
& $dx_2$ & $-(v_1+v_3)/(1-v_2)$ \\
\hline
$\Omega \varepsilon \Lambda \alpha o$ &  & $dx_1-dx_2$
& $-dx_3$ & $(v_1-v_2)/(1+v_3)$ \\
\hline
 &$\Delta \beta \Phi \delta o$ & $-(dx_1-dx_2)$
& $-dx_3$ & $-(v_1-v_2)/(1+v_3)$ \\
\hline
$\Omega \alpha \Upsilon \zeta o$ &  & $dx_1-dx_3$
& $-dx_2$ & $(v_1-v_3)/(1+v_2)$ \\
\hline
 &$\Delta \delta \Theta \gamma o$ & $-(dx_1-dx_3)$
& $-dx_2$ & $-(v_1-v_3)/(1+v_2)$ \\
\hline
$\Theta \gamma \Lambda \varepsilon o$ &  & $dx_3-dx_2$
& $-dx_1$ & $(v_3-v_2)/(1+v_1)$ \\
\hline
 &$\Upsilon \zeta \Phi \beta o$ & $-(dx_3-dx_2)$
& $-dx_1$ & $-(v_3-v_2)/(1+v_1)$ \\
\hline
\end{tabular}
\end{center}

\end{document}